\documentstyle[11pt,newpasp,twoside,epsf]{article}
\newcommand{\sm}{\mbox{M}_{\odot}}
\newcommand{\be}{\begin{equation}}
\newcommand{\ee}{\end{equation}}
\newcommand{\te}{t_{\rm e}}
\newcommand{\tfw}{t_{\rm FWHM}}
\newcommand{\tst}{\textstyle}

\newcommand{\gs}{\;\raisebox{-.8ex}{$\buildrel{\textstyle>}\over\sim$}\;}

\def\edcomment#1{\iffalse\marginpar{\raggedright\sl#1\/}\else\relax\fi}
\reversemarginpar

\begin{document}
\title{Pixel lensing towards M31 in principle and in practice}
\author{Eamonn Kerins for the POINT-AGAPE Collaboration}
\affil{Theoretical Physics, University of Oxford, \\ 1 Keble Road, Oxford
OX1 3NP, UK}

\begin{abstract}
The Andromeda galaxy (M31) provides a new line of sight for Galactic
MACHO studies and also a signature, near-far asymmetry, which may
establish the existence of MACHOs in the M31 halo. We
outline the principles behind the so-called ``pixel-lensing''
experiments monitoring unresolved stars in M31. We present detailed
simulations of the POINT-AGAPE survey now underway, which is using the
INT wide-field camera to map the microlensing distribution over a
substantial fraction of the M31 disc. We address the extent to which
pixel-lensing observables, which differ from those of classical
microlensing, can be used to determine the contribution and mass of
M31 MACHOs. We also present some preliminary light-curves obtained from
the first season (1999/2000) of INT data.
\end{abstract}

\section{Andromeda: the new frontier}

This meeting is in part a celebration of the success of the current
pioneering generation of microlensing surveys, the discoveries of
which are reported throughout these proceedings. However, as with all
pioneering work, a number of puzzles remain to be solved including,
crucially, whether MACHOs exist or whether current results towards the
Large and Small Magellanic Clouds (LMC and SMC) can be explained by
ordinary stellar populations.  Whilst the MACHO Collaboration
interprets the excess of events it finds as evidence of a $\sim 20\%$
halo contribution in objects of $\sim 0.5~\sm$ (Alcock et al. 2000)
the EROS Collaboration has chosen to place only upper limits on the
halo fraction from its sample of LMC/SMC events, even though the EROS
dataset is formally consistent with that of MACHO (Lasserre et
al. 2000). This dichotomy in approach reflects a community-wide
uncertainty as to the contribution to the microlensing rate from the
Magellanic Clouds themselves. As the current LMC/SMC surveys come to
an end it is possible that the Magellanic satellite galaxies will
continue to represent clouds of uncertainty for MACHOs.

Crotts (1992) and Baillon et al. (1993) proposed searching for MACHOs
towards the Andromeda galaxy (M31). There are several aspects which
make M31 very appealing for microlensing studies. Firstly, it is a
large external galaxy, with ${\cal O}(10^{10})$ sources, and has a
halo of its own, providing an additional site for MACHOs. We should
therefore expect many more events than towards the Magellanic
Clouds. Small number statistics ought not be an issue for M31 if the
MACHO abundance is significant. Secondly, the additional line of sight
through our own halo that M31 provides may help to decide the
viability of some of the stellar self-lensing scenarios, which have
been proposed as non-MACHO alternatives to explain the LMC/SMC
events. Alcock et al. (1995) have also noted that the ratio of
microlensing rates between M31 and the LMC could be used in principle
to probe the outer rotation curve of our Galaxy, though in practice
the background from events in M31 makes this difficult. Lastly, our
external viewpoint is advantageous in that we can map the MACHO
distribution across the face of the M31 disc. The importance of this
has been emphasized by Crotts (1992), who noted that the $77\deg$
inclination of the M31 disc should give rise to a noticeable gradient
in the microlensing rate if MACHOs occupy a spheroidal distribution,
as depicted in Figure~1. Such an asymmetry does not arise naturally
for disc lensing or variable stars, and so would provide strong
evidence for the existence of MACHOs if detected.
\begin{figure}
\begin{minipage}{6.5cm}
\plotfiddle{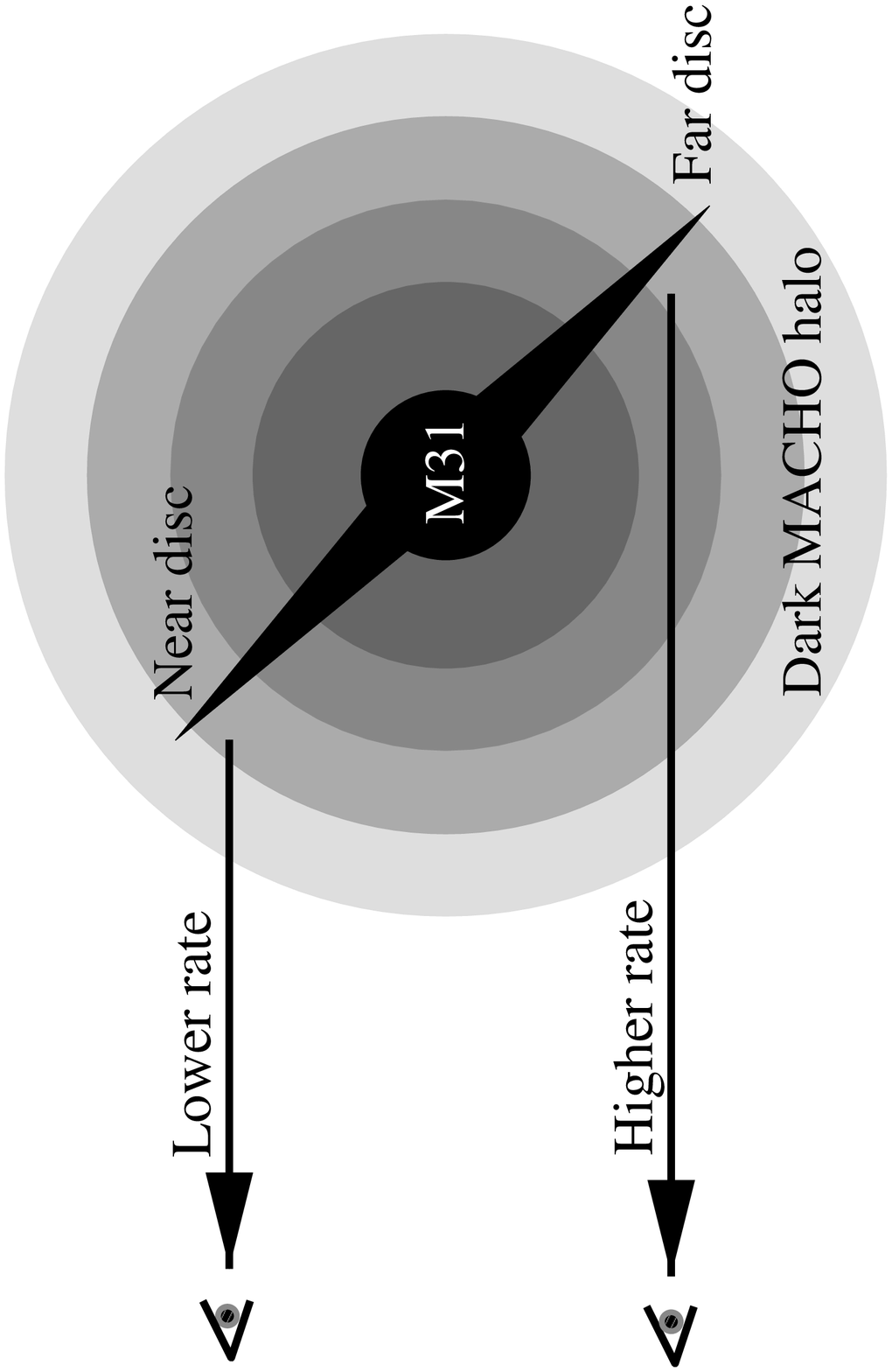}{0cm}{270}{30}{30}{80}{-20}
\caption{The concept of near-far asymmetry. Due to the tilt of the M31
disc the microlensing rate is larger towards the far disc than towards the
near side if MACHOs are present in a spheroidal dark halo.  The effect
is less pronounced if the halo is flattened. The distributions of disc
events and variable stars are symmetric.}
\end{minipage}
\end{figure}

\section{Pixel lensing}

There are a number of difficulties in conducting a ground-based
programme towards more distant targets like Andromeda. The major
problem is that, in general, sources are resolved only whilst being
lensed. Towards the M31 bulge the stellar surface densities may reach
several thousand per arcsec$^2$, so after the event one often cannot
identify the source to measure its baseline flux. In practice one
must monitor pixel flux rather than individual source flux,
hence the term {\em pixel lensing}\/.  Because of the pixel flux
contribution of the unlensed sources one typically requires the lensed
source to be intrinsically bright or highly magnified, so the majority
of microlensing events escape detection. In the high-magnification
regime the light-curve suffers from a well-known near-degeneracy
in which the excess flux due to microlensing is
   \be
      \Delta F (t) \simeq \frac{A(t_0) F_{\rm s}}{\sqrt{1+\left[
      \frac{\tst t-t_0}{\tst \te A(t_0)^{-1}} \right]^2 }} \quad \quad
      \quad \quad (A \gs 10),
   \ee
where $t$ denotes observation epoch, $t_0$ is the epoch at which the
magnification, $A$, attains its maximum, $F_{\rm s}$ is the baseline
source flux and $\te$ the Einstein radius crossing time. As Wo\`zniak
\& Paczy\`nski (1997) have pointed out, equation~(1) is invariant
under transformations $F_{\rm s} \rightarrow F_{\rm s} / \alpha$, $A
\rightarrow \alpha A$, and $\te \rightarrow \alpha \te$ for
constants $\alpha$ preserving the high-magnification regime. Under
such circumstances one is unable to unambiguously determine $\te$, an
important observable for conventional microlensing studies. All of
this is made worse still by variations in seeing, sky background,
detector position and point spread function which make it difficult to
isolate the microlensing signal. Changes in seeing pose a severe
problem because they induce localized flux variations on
timescales comparable to microlensing.

These challenges, though difficult, have been met. Two pilot
experiments able to detect true source flux variation on a routine
basis have demonstrated the technical viability of pixel lensing
(Crotts \& Tomaney 1996; Ansari et al. 1997). The experiments employed
different techniques, difference imaging and superpixel photometry, to
compensate for changes in seeing. However, the limiting factor for both
surveys has been the small field of view, yielding relatively
modest statistics.

\section{POINT-AGAPE}

\begin{figure}
\plotfiddle{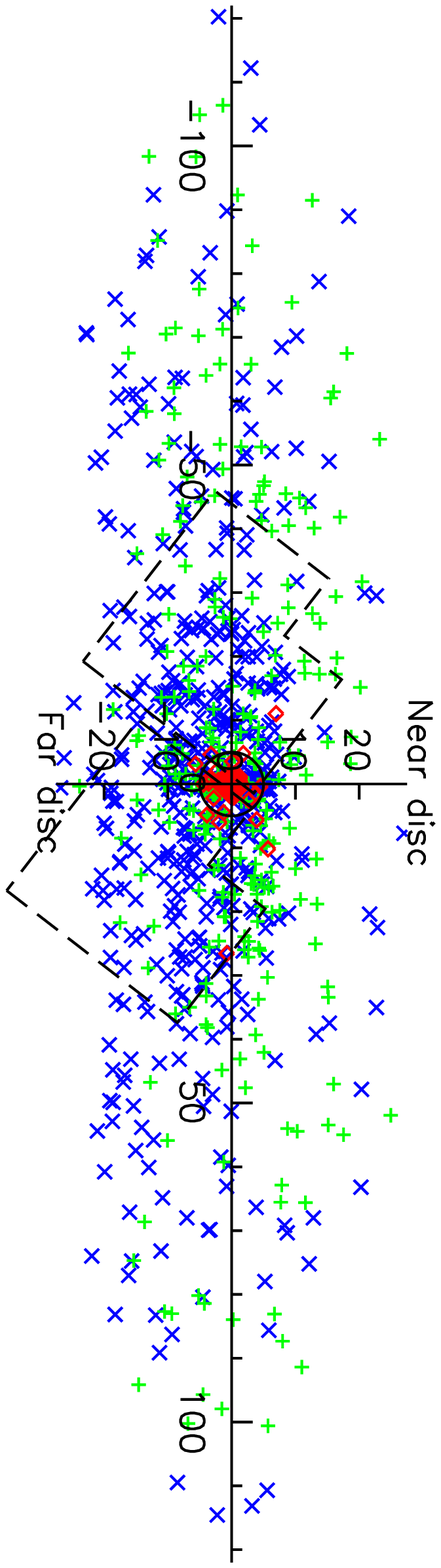}{2.9cm}{90}{50}{50}{200}{-160}
\plotfiddle{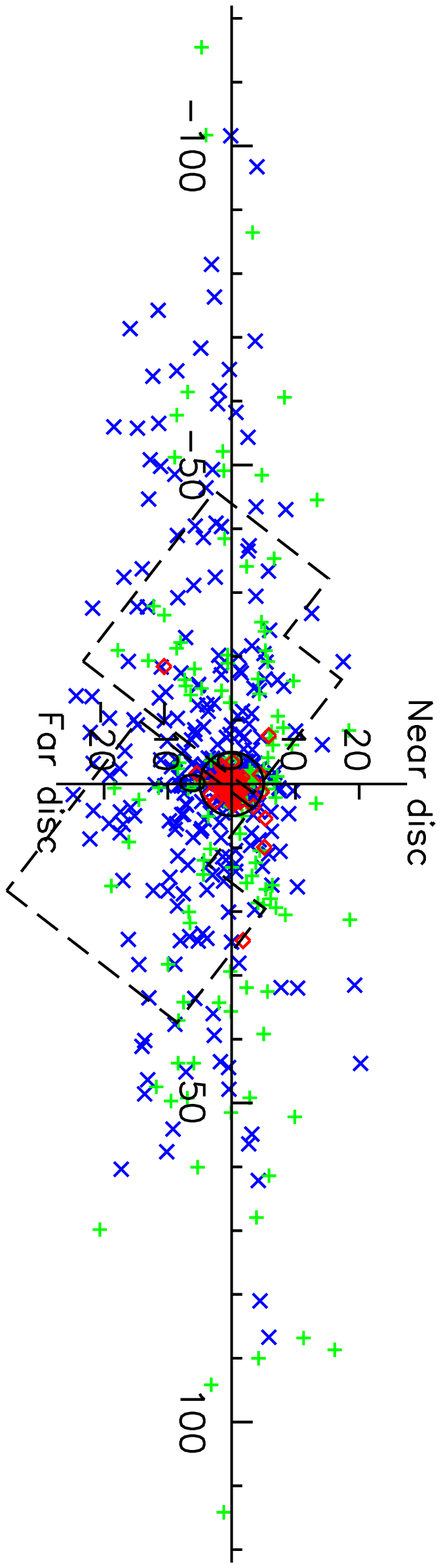}{2.9cm}{90}{50}{50}{200}{-160}
\plotfiddle{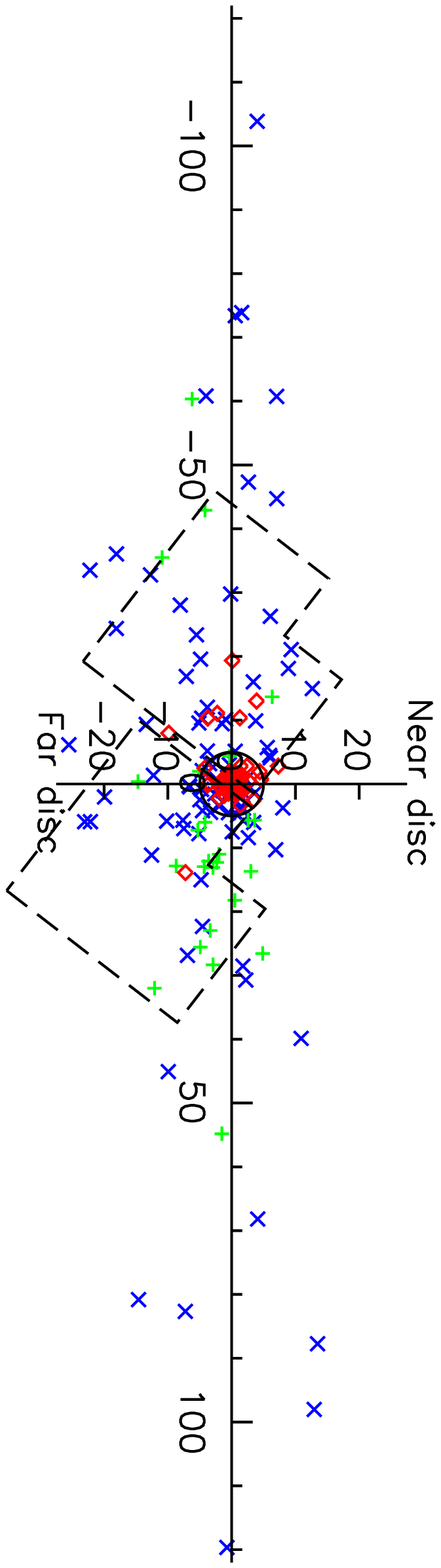}{2.9cm}{90}{50}{50}{200}{-160}
\caption{The simulated spatial distribution of pixel-lensing events
for the POINT-AGAPE survey after three observing seasons. The axes are
in arcmins. Galaxy MACHOs are shown as crosses, M31 MACHOs are depicted
by an ``x'' and stellar events by diamonds. We assume here that MACHOs
have a mass of 0.1 ({\em top}\/), 1 ({\em middle}\/) and $10~\sm$
({\em bottom}\/) and provide all the dark matter in the Galaxy and M31
haloes. The dashed lines indicate the POINT-AGAPE fields whilst the
inner circle denotes the central region dominated by stellar
self-lensing.}
\end{figure}

Building upon the pioneering French AGAPE (Andromeda Galaxy Amplified
Pixels Experiment) programme, the Anglo-French POINT-AGAPE survey is
employing the INT wide-field camera (WFC) to image a large fraction of
the M31 disc (POINT is an acronym for Pixel-lensing Observations with
INT). The survey began in August 1999 and is undertaking multi-colour
observations in two fields, covering 0.6~deg$^2$ of the disc. The
first season of data collection was completed in January 2000.

We have performed detailed simulations to model the sensitivity of the
POINT-AGAPE survey, incorporating the effects of seeing and variable sky
background, as well as our sampling (Kerins et al. 2000). The
simulations include spherical near-isothermal models for both the
Galaxy and M31 haloes, as well as a sech-squared disc and exponential
bulge for the M31 stellar components. For the haloes we have simulated
the expected signal for a wide range of MACHO masses, whilst for the
stellar components we have assumed a Solar neighbourhood mass
function. 

Simulations of the expected spatial distribution of observed events
after three seasons for 0.1, 1 and $10~\sm$ MACHOs are provided in
Figure~2. The figure assumes the Galaxy and M31 haloes are full of
MACHOs. The POINT-AGAPE fields are shown by the dashed-line
templates. Whilst the fields cover about one-fifth of the area of the
disc their central location means that about half of all detectable
pixel events across the disc should occur within them. The high
concentration of events within the inner 5~arcmin is due mostly to
bulge-bulge lensing, where as the MACHOs are dispersed over the entire
disc. The near-far asymmetry in M31 MACHOs gives rise to a overall
asymmetry in the spatial distribution. One can see from Figure~2
how the number of MACHOs increases and the near-far asymmetry becomes
more prominent at small mass scales

\begin{figure}
\begin{minipage}{7.cm}
\plotfiddle{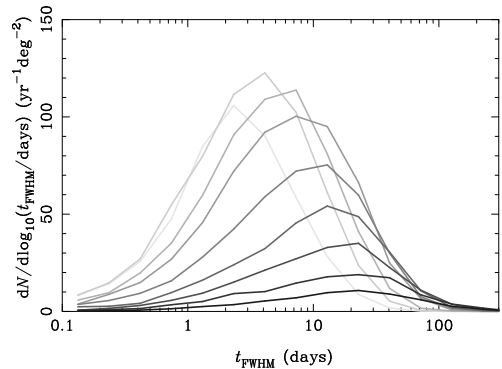}{0.cm}{270}{28}{28}{80}{20}
\caption{The combined M31 and Galaxy MACHO $\tfw$ rate distributions
for a range of MACHO masses, normalized to full MACHO haloes and
averaged over the M31 disc. From the lightest to the darkest curve the
Galaxy and M31 MACHO mass is 0.001, 0.003, 0.01, 0.03, 0.1, 0.3, 1, 3
and $10~\sm$.}
\end{minipage}
\end{figure}
Since the Einstein duration $\te$ is usually not measurable we have
computed the expected event timescale distribution in terms of the
full-width at half-maximum flux $\tfw$. Figure~3 displays the
normalized combined Galaxy and M31 MACHO timescale distribution for a
range of mass scales, assuming a maximal MACHO contribution and
averaging the rate over the M31 disc. The distributions indicate that
the average duration $\langle \tfw \rangle$ is sensitive to MACHO
mass, though for our sampling it is much less sensitive than the
underlying average Einstein duration $\langle \te \rangle$, with
$\langle \tfw \rangle \propto \langle \te \rangle^{1/2}$. The rate
peaks for $0.003-0.01~\sm$ MACHOs, giving $\sim 100$ events per season
for M31 MACHOs and $\sim 40$ per season for Galaxy MACHOs within the
two POINT-AGAPE fields if both haloes are full of MACHOs. If we
instead use the most recent Galaxy MACHO mass estimate of $0.5~\sm$,
with a $20\%$ halo contribution, then we should expect about 15 MACHO
detections per season.

\begin{figure}
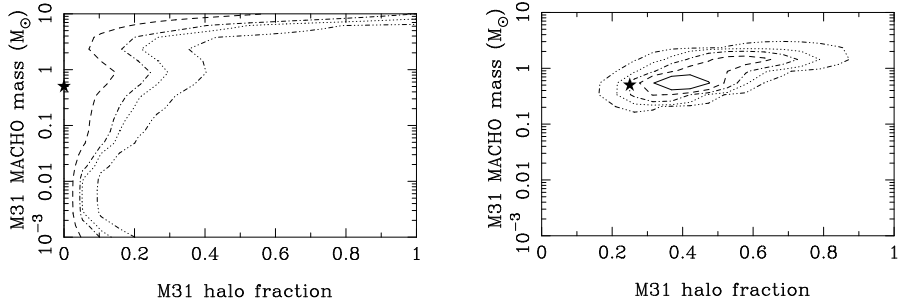

\plotfiddle{kerins4a.ps}{4.cm}{270}{25}{25}{-180}{120}
\plotfiddle{kerins4b.ps}{0.cm}{270}{25}{25}{0}{145}
\caption{Simulated maximum-likelihood recovery of M31 MACHO
parameters. The {\em left panel}\/ shows the first-season
sensitivity of POINT-AGAPE to an absence of near-far asymmetry, and
hence an absence of M31 MACHOs (only stellar lensing events and
variable stars have been generated in this simulation). In the {\em
right panel}\/ we assume a MACHO mass and contribution as indicated by
the star. The contours here illustrate the survey sensitivity after three
seasons. For both plots the solid, dashed, dot-dashed, dotted and
triple-dot-dashed contours indicate 34\%, 68\%, 90\%, 95\% and 99\%
confidence levels, respectively.}
\end{figure}
We have assessed the extent to which the MACHO mass and halo fraction
can be recovered from the observables, specifically the event location
and $\tfw$. Artificial datasets were constructed from our simulations
and a maximum likelihood estimation of MACHO parameters computed using
a Bayesian likelihood estimator. As well as M31 MACHOs the datasets
include a fixed contribution from stellar lensing and may include
Galaxy MACHOs with a different mass and density contribution to the
M31 MACHOs. We also make allowances for dataset contamination due to
variable stars passing our microlensing selection criteria. Our
likelihood estimator is therefore computed over a five-dimensional
likelihood space covering M31 MACHO mass and density; Galaxy MACHO
mass and density; and a variable star contamination level. In Figure~4
we have summed the likelihood over three of the five dimensions to
highlight the sensitivity of POINT-AGAPE to the M31 MACHO mass and
halo fraction. The stars in each panel indicate the input parameters
used to generate the artificial dataset whilst the contours indicate
the parameter recovery from the dataset. In the left-hand panel our input
MACHO fraction is zero, only stellar lensing events and variable stars
have been generated. The contours indicate the POINT-AGAPE sensitivity
after just one season. Powerful constraints are obtained over a wide
range of mass scales due to the lack of any near-far asymmetry. The
right-hand panel shows the sensitivity after three seasons for
currently favoured MACHO parameters. The constraints on M31 parameters
achievable by POINT-AGAPE are comparable to those being obtained for
Galaxy MACHOs by the LMC/SMC surveys. 

\begin{figure}
\begin{minipage}{6.5cm}
\plotfiddle{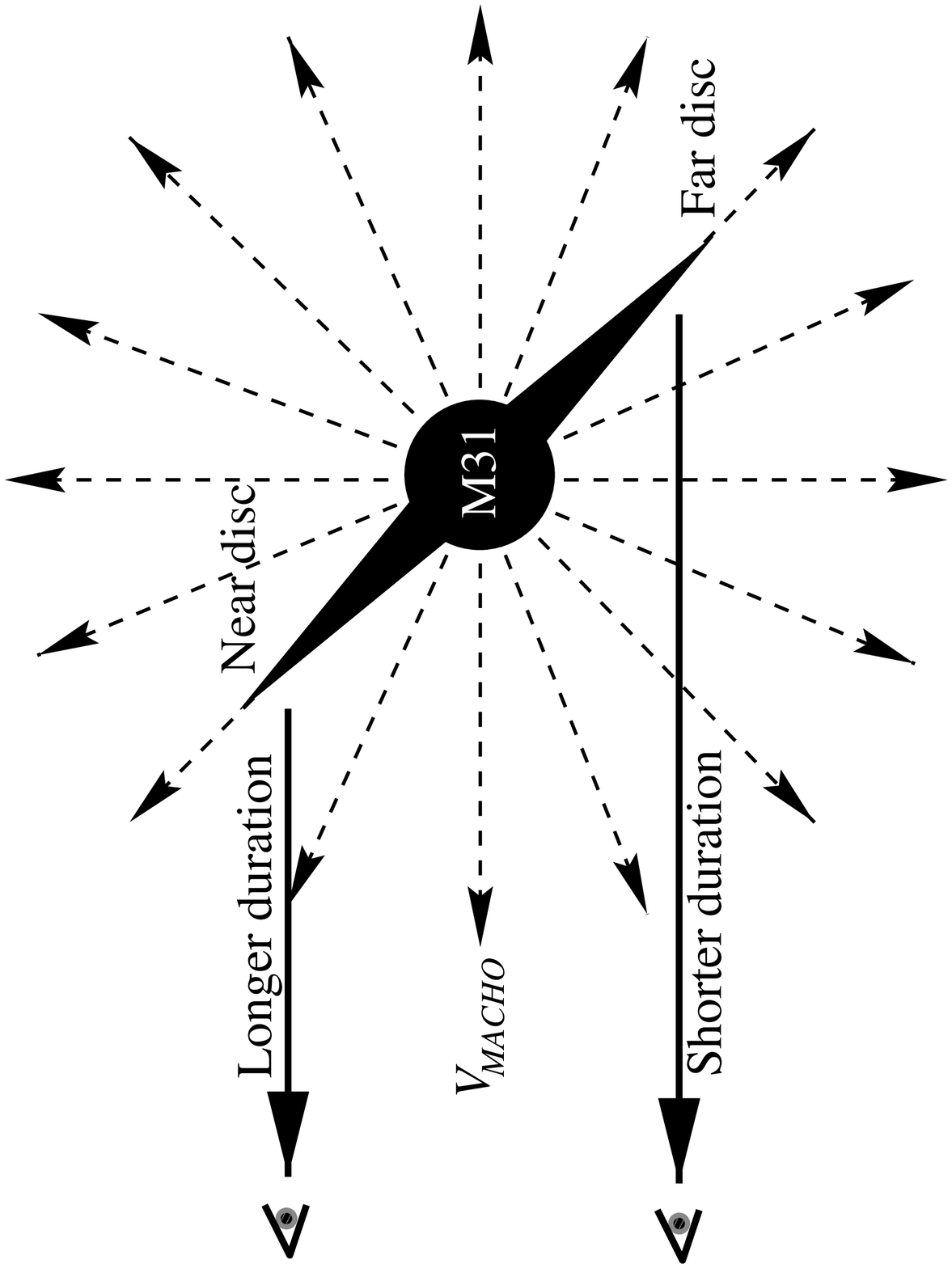}{0cm}{270}{30}{30}{80}{-10}
\caption{Dual spatial and timescale asymmetry as a probe of velocity
anisotropy. If MACHOs have strongly radial orbits then their motion
will tend to be parallel to the near-disc sight-line but orthogonal to
the far-disc sight-line. Hence near-side events should last longer and
have a lower rate, enhancing the spatial asymmetry over that expected
for isotropic haloes.}
\end{minipage}
\end{figure}
Pixel-lensing surveys towards M31 have the potential not just to probe
the halo contribution and mass of MACHOs, but also their distribution
function. The near-far asymmetry effect is sensitive to halo
flattening, though for highly flattened haloes the asymmetry
disappears, making it difficult to distinguish between M31 and Galaxy
MACHOs, not to mention variable stars. In Figure~5 we illustrate how
near-far asymmetry could also be used to probe the velocity anisotropy
of M31 MACHOs. If the MACHOs have strongly radial orbits their
trajectories should cut near-side and far-side lines of sight at
different angles, giving rise to different characteristic event
durations. Specifically, event durations should be longer for the
near-side and the rate should be correspondingly reduced (since, for a
given optical depth, the rate $\Gamma \propto \langle \te
\rangle^{-1}$) relative to isotropic halo models. In this case we
would observe two signatures: an enhanced near-far spatial asymmetry
and a near-far timescale asymmetry. This combination of signatures
would be difficult to arrange by other means and would therefore
represent a strong argument for MACHOs and for an anisotropic MACHO
distribution function.

Analysis of the 1999/2000 INT dataset is yet to be finalized, but a
preliminary reduction reveals many interesting light-curves, including
a variety of variable stars and possible microlensing candidates. Some
sample light-curves, spanning about a third of the 1999/2000 baseline,
are shown in Figure~6. The light-curves have been processed using
superpixel photometry, which is described in detail in Ansari et
al. (1997). Briefly, the flux in each pixel is firstly represented by
the summed flux over a pixel array, or superpixel, centred on the
pixel. The size of the array is set by the size of the pixel relative
to the seeing scale, and in the case of the INT WFC a $7\times 7$
superpixel array was chosen. Though this binning dilutes any
microlensing signal that may be present it helps to stabilize the
effects of seeing variations to a point where a simple linear
correction can be applied to the superpixel flux on each image in
order to match their seeing characteristics. The method is simple and
yet permits differential photometry practically down to the photon
noise limit.

\begin{figure}
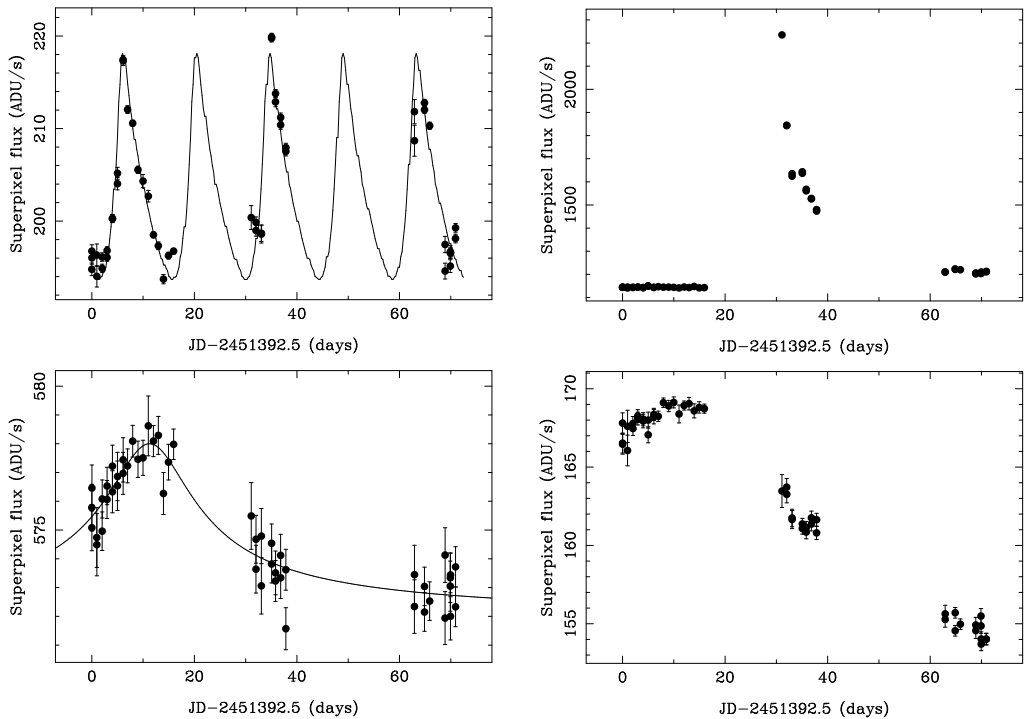

\plotfiddle{kerins6a.ps}{3.5cm}{270}{28}{28}{-200}{130}
\plotfiddle{kerins6b.ps}{0.cm}{270}{28}{28}{0}{154}
\plotfiddle{kerins6c.ps}{3.5cm}{270}{28}{28}{-200}{130}
\plotfiddle{kerins6d.ps}{0.cm}{270}{28}{28}{0}{154}
\caption{Preliminary $r$-band light-curves spanning one-third of the
1999/2000 POINT-AGAPE baseline. {\em Top left}\/: a possible Cepheid,
together with a model fit. {\em Top right}\/: a probable nova. {\em
Bottom left}\/: a light-curve consistent with microlensing, together
with a model fit with $\tfw = 28$~days. {\em Bottom right}\/: another
possible microlensing event. A longer baseline is required before
either light-curve could be classified as a microlensing candidate.}
\end{figure}
The top-left and top-right light-curves in Figure~6 are consistent
with a Cepheid and a nova, respectively. They illustrate the high
signal-to-noise ratio achievable with INT WFC superpixel
photometry. The bottom panels show light-curves consistent with a
microlensing interpretation. The light-curve at bottom-left peaks at
about $r \simeq 22$~mag and has been fit with a theoretical degenerate
microlensing curve with $\tfw = 28$~days for illustration. A longer
baseline is required before either light-curve could be flagged as a
microlensing candidate since long-period variable stars, such as Miras,
pose a serious problem for pixel-lensing selection. To reject such
variables one requires at least a three-year baseline. In the short
term one can use the presence or absence of near-far asymmetry to
statistically measure candidate contamination, since variable stars
should have a symmetric spatial distribution.

There are of course limitations to pixel lensing. In common with
Galactic microlensing, pixel-lensing constraints are sensitive to the
assumed halo distribution function and, as already mentioned, the
reduced near-far asymmetry for flattened haloes means that M31 MACHOs
in such a configuration will be harder to detect or constrain. The
pixel-lensing predictions are additionally sensitive to the shape of
the bright end of the M31 stellar luminosity function, which is still
to be properly determined. Near-far asymmetry nonetheless remains a
very powerful discriminant between MACHOs and stellar self-lensing, an
issue which currently hampers the interpretation of the LMC/SMC
events. As well as searching for MACHOs, we also hope to be able to
constrain the structure and mass function of the M31 bulge, where many
self-lensing events are expected to occur. We are optimistic that,
within about three seasons, the POINT-AGAPE survey will quantify and
constrain both the stellar and MACHO mass function and distribution in
M31, furthering our knowledge of the nature of dark matter, and
providing a powerful probe of galactic structure.

\end{document}